\documentclass[10pt,letterpaper,twocolumn]{article}
\usepackage{tipa}
\usepackage{amsfonts}
\usepackage{amssymb} 

\usepackage{ol2}
\usepackage[draft]{hyperref}
\usepackage{amsmath}

\begin{document}

\twocolumn[ 

\title{Sub-50 attosecond pulse generation from multicycle nonlinear chirped pulses}


\author{Yang Xiang,$^{1,2}$ Yueping Niu,$^{1,*}$ Yihong Qi,$^1$ and Shangqing Gong $^{1,\dagger }$}

\address{
$^1$State Key Laboratory of High Field Laser Physics, Shanghai
Institute of Optics and Fine Mechanics, Chinese Academy of Science,
Shanghai 201800, China
\\
$^2$College of Computer Science and Technology, Henan Polytechnic University, Jiaozuo 454000, China \\

Corresponding author: $^*$ niuyp@siom.ac.cn, $^\dagger$
sqgong@siom.ac.cn }

\begin{abstract}We present a method of producing single attosecond
pulses by high order harmonic generation with multi-cycle nonlinear
chirped driver laser pulses. The symmetry of the laser field in
several optical
 cycles near the pulse center is dramatically broken, and then the photons which cover a much broad spectrum burst almost only
 in one optical cycle. So an ultra-broad continuum spectrum appears in the high order harmonic spectrum,
 from which an isolated sub-50 attosecond pulse could be obtained.
 The results are almost independent of the length and chirp form of the driver laser pulse.\end{abstract}

]  

\noindent The creation of attosecond(as) pulses has opened the door
to the new field which is called attosecond science. Attosecond
pulses are of great importance for studying and controlling the
coherent dynamics of electrons on their natural time scales. So far,
attosecond pulses are mainly obtained from the high harmonic
generation (HHG) process$^{1-7}$. By selecting many harmonics in the
plateau which only contains odd harmonics of the laser frequency,
one can obtain an attosecond train. This was observed by Paul
\textit{at al} in experiments for the case of multicycle driver
pulses$^1$. Single attosecond pulse can be generated from the
continuous spectrum near the cutoff of the HHG from few-cycle laser
pulse. Up to now, the shortest pulse duration achieved by this means
is about 80 attoseconds$^2$. While the pulse duration used in
experiment is shorter than 5 fs, which is so stringent that it can
be achieved only by means of state-of-the-art laser technology. So,
the single attosecond obtained from multicycle laser pulses attracts
great interests. Several schemes have been proposed for the
generation of single attosecond in multicycle-driver regime, such as
polarization gating technique$^3$, two-color control$^4$ and
waveform control$^5$. While the durations of the pulses obtained by
these methods are about several hundred attoseconds, though Hong
\textit{et al} pointed out that a sub-100 isolated attosecond pulse
could be obtained by two-color control$^6$, the duration of the
laser pulse (800nm) used is 10 fs, which is still shorter than the
typical performance of standard laser system (${>}$ 25 fs)$^3$.

In this Letter, we propose an approach to achieve a single
attosecond pulse from nonlinear chirped multicycle driving pulses.
The prominent advantage of this approach is that the duration of the
pulse is less than 50 attoseconds, which is much shorter than that
obtained from the other method in multicycle-driver regime, and the
result is almost independent of the duration of the laser pulse and
the chirp form. The physical mechanism of the single attosecond
generation is connected with the well-known semi-classical three
step model$^8$ (TSM): first, the electron tunnels through the
barrier formed by the Coulomb potential and the laser field; next,
it oscillates almost freely in the laser field; finally, it may
return back and recombine with the parent ion at later time. During
the recombination, a photon is emitted. Usually, all of this occurs
in every half-cycle which may cause the generation of attosecond
pulse train. If the electron has few collisions with the nuclear,
then the HHG spectrum become continuous, from which a single
attosecond pulse can be obtained$^9$. In the following, we first use
a classical calculation based on the TSM to explore the physical
origin of the single attosecond pulse genration. Afterwards, we
perform a quantum mechanical simulation using a one-dimensional
helium atom model to compare with the traditional method, by solving
the time-dependent Schr\"odinger equation(TDSE). Then we explore the
impact of the duration and chirp form on the generation of single
attosecond pulse. At last, the conclusion is given.

At first, we consider the classical motion of an electron in a
linearly polarized chirped multicycle laser pulse, which has a form
of [the atom units (a.u.) are used in all equations in this paper,
unless otherwise mentioned.]:

\begin{align}
{{E}\left(t\right)} &={F}\sin^{2}\left(
\frac{\pi{t}}{\tau}+\frac{\pi}{2}\right)\cos\left[\omega{t}+\delta\left({t}\right)\right],
\end{align}
where \textit{F}, $\omega$ and $\tau$  is the amplitude, frequency
and the pulse length of the laser field. The time-dependent carrier
envelop phase (CEP) is set to be$^{10}$:
\begin{align}
{{\delta}\left(t\right)}
&=-\beta\tanh\left(\frac{t-t_0}{\sigma}\right).
\end{align}
The parameters $\beta$, \textit{t$_0$} and $\sigma$ are used to
control the chirp form. Such a time-varying CEP can be achieved by
means of the comb laser technology$^{10-12}$. In our calculations,
\textit{F} is 0.12 a.u. (corresponds to intensity \textit{I}=
5.0$\times$10$^{14}$ W/cm$^2$)and $\omega$ is 0.057 a.u.
(corresponds to wavelength $\lambda$=800 nm). The chirp parameters
are as follows: $\beta$=8.0, $\sigma$=200 a.u. and
t$_0$=$\sigma$/4.0. According to the TSM, the return energy of the
electron can be obtained by solving the Newton equation
\textit{dv}/\textit{dt}=-\textit{E}(\textit{t}), where \textit{v}
represents the velocity of the electron. The return energies of the
electron are shown in Fig.1, as well as the time evolution of the
chirped laser pulse whose length is 60 optical cycles (O.C.)[the
full width at half maximum (FWHM) is about 58 fs]. From the return
energy map, we can see that the max return energy extends to about
30\textit{U$_p$}, which is much larger than the well known value
3.17\textit{U$_p$}$^{13}$, where \textit{U$_p$}
=\textit{F}/(4$\omega$$^2$), is the ponderomotive energy in the
laser field. This is due to the great asymmetry of the laser field
near the pulse center, which is discussed in detail in our previous
work$^{14}$. Moreover, the electrons with energy larger than about
11\textit{U$_p$} only return within less than 1 O.C., as shown in
Fig.1(b). This means that the photons which cover a much broad
spectrum burst only in less than 1 O.C., so there should be a much
broad continuum spectrum in the HHG, from which a single attosecond
pulse could be obtained.

\begin{figure}[htb]
\centerline{\includegraphics[width=8cm,height=4cm]{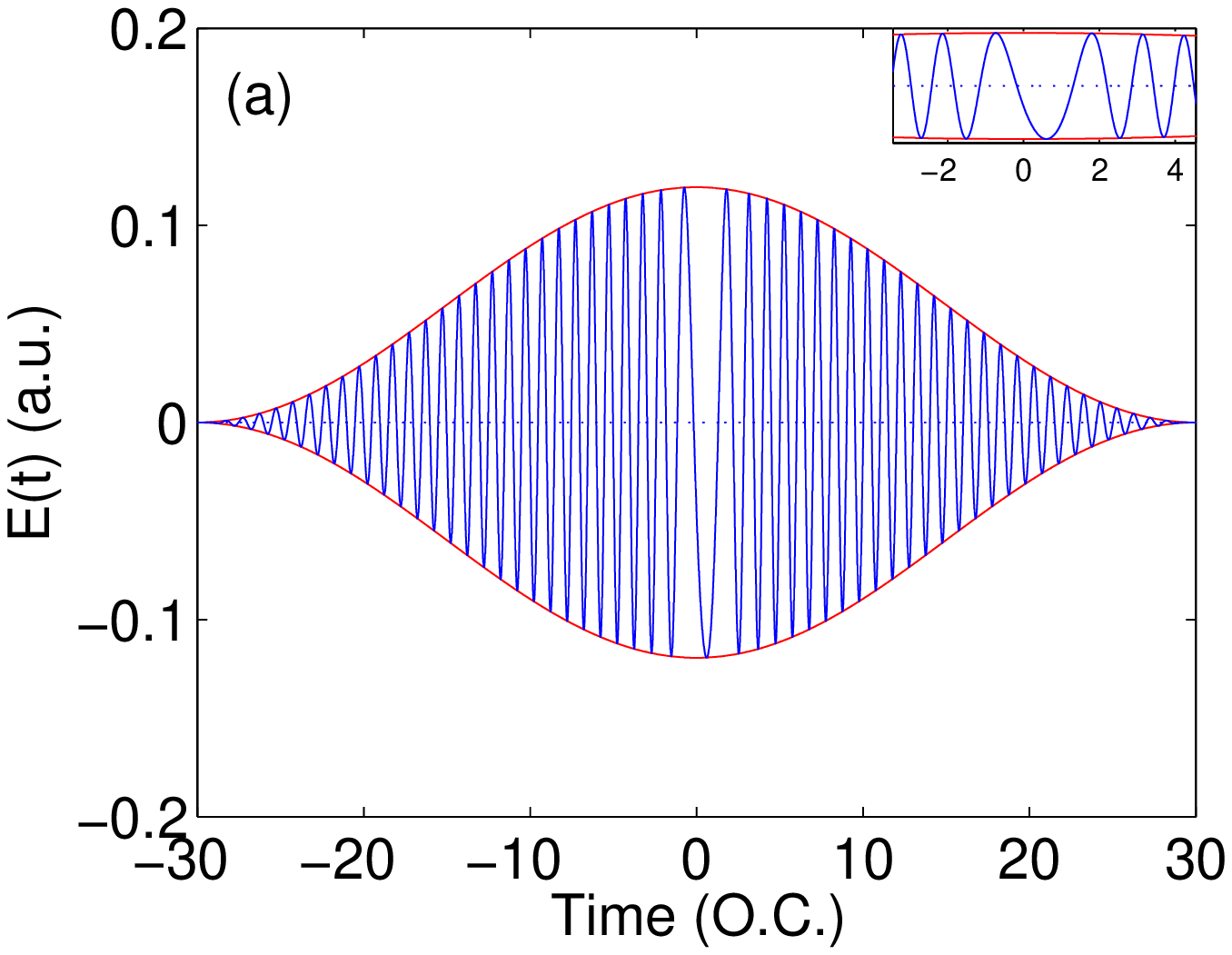}}
\centerline{\includegraphics[width=8cm,height=4cm]{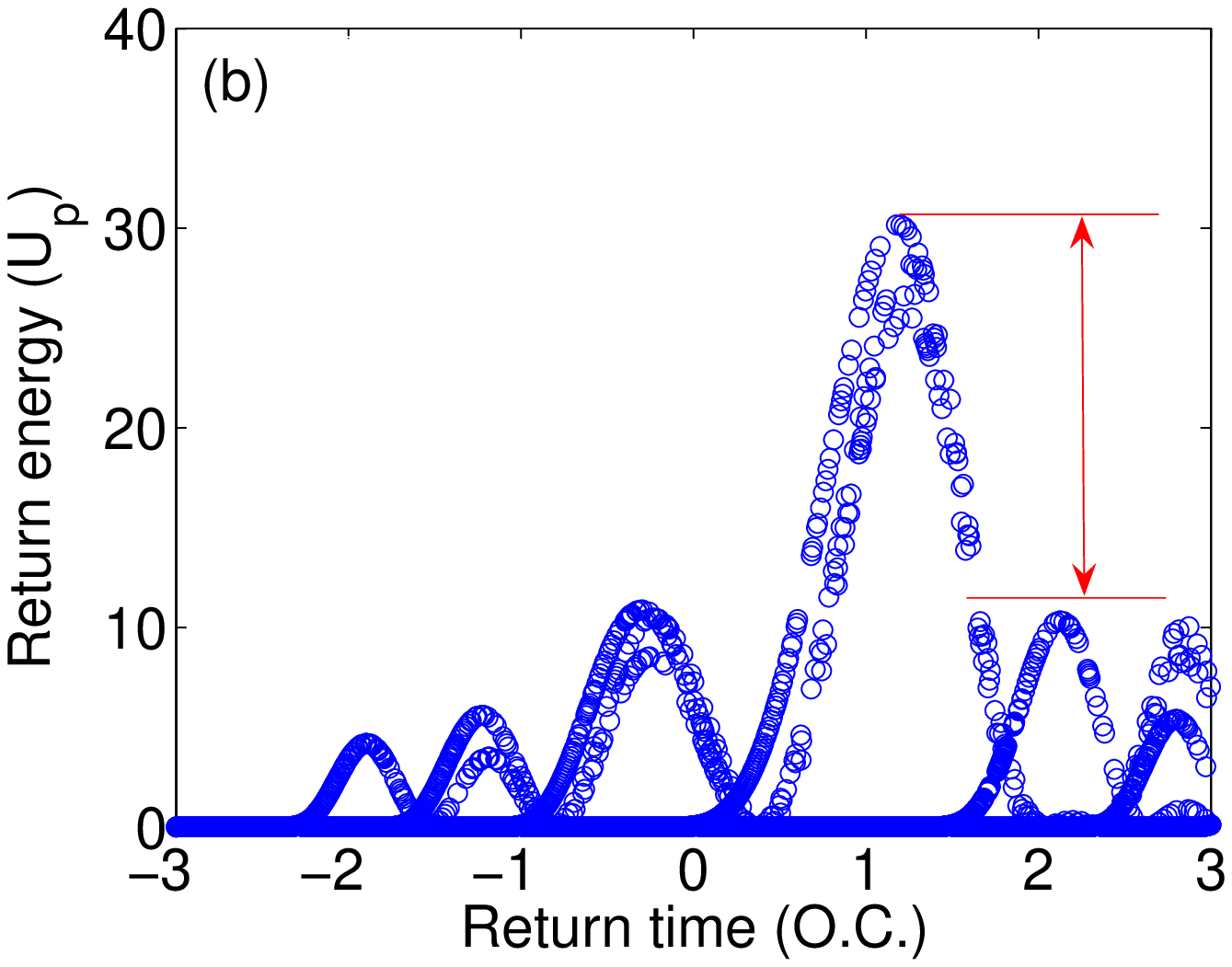}}
\caption{(Color online) (a) The time evolution of the chirped laser
pulse and its envelope. The inset window is the enlargement electric
field near the center of the pulse. (b) Classical return energy
map.}
\end{figure}
In order to verify the classical analysis above, we perform a
simulation of high harmonic attosecond pulse generation by solving
the TDSE. Our model system is a one-dimensional helium atom with a
single active electron$^{15}$ and dipole approximation. The method
for solving the one-dimensional TDSE is split-operator
method$^{16,17}$. Once the wavefunction is obtained, the HHG
spectrum can be obtained by taking the Fourier transform of the
dipole acceleration, applying the Ehrenfest's theroem$^{18}$. The
HHG spectrum for the laser field in Fig.1(a) is shown in Fig.2(red
solid line). We can see that the HHG spectrum exhibits two plateaus:
the cutoff of the first one is about at the 230th order harmonic,
corresponding to the energy \textit{I$_p$}+11\textit{U$_p$}, where
\textit{I$_p$}(=24.6 eV) is the ionization potential of helium, and
the cutoff of the second plateau is about at the 594th order
harmonic, corresponding to the energy
\textit{I$_p$}+30\textit{U$_p$}. Compared with the return energy map
of Fig.1(b), the result obtained from quantum theory is quite in
agreement with that from the classical method.

\begin{figure}[htb]
\centerline{\includegraphics[width=8cm,height=4cm]{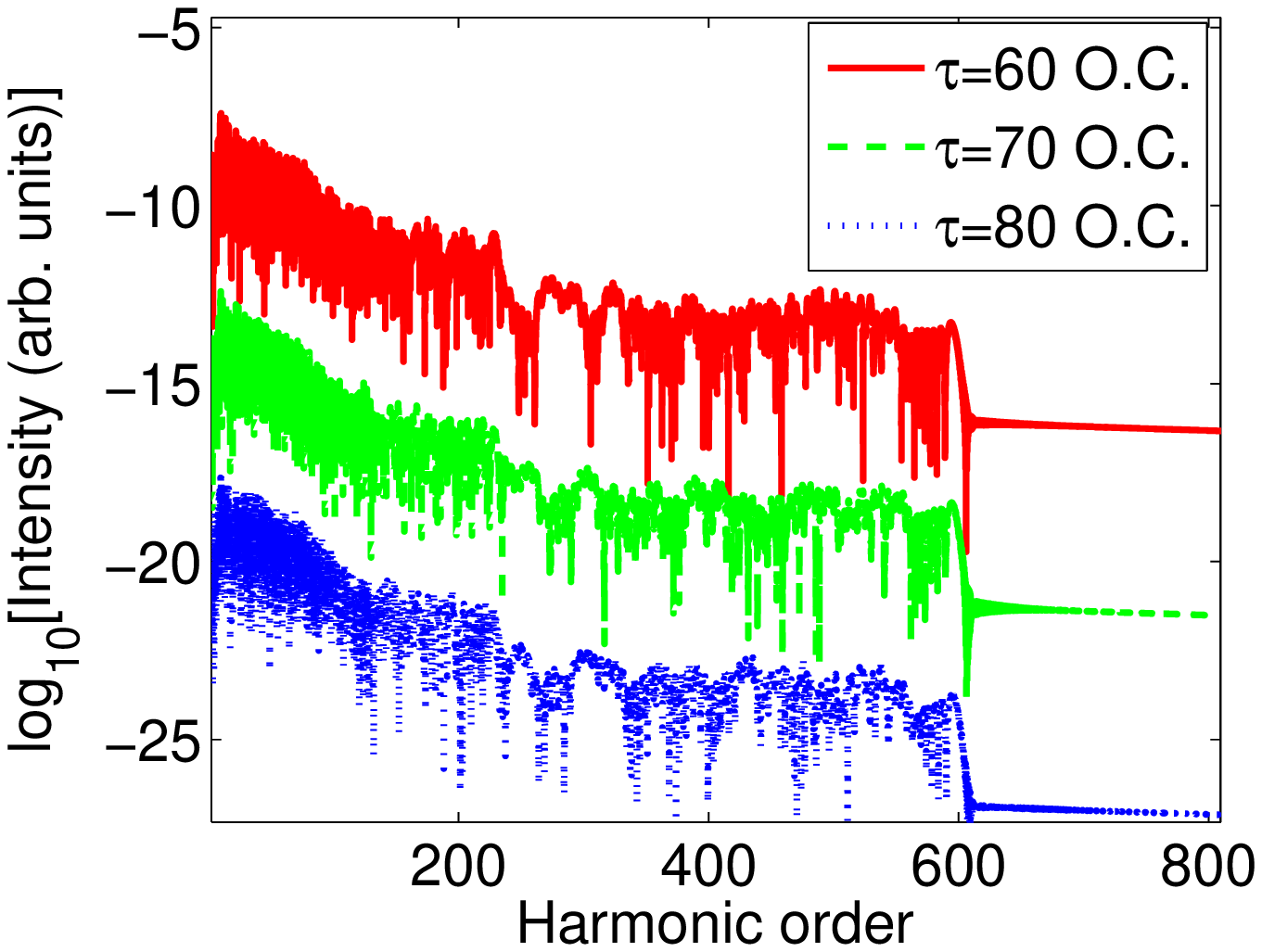}}
\caption{(Color online) The HHG spectrums of the chirped laser
pulses with different lengths. For clarity, the curves have been
offset along the vertical axis.}
\end{figure}

Now we consider the attosecond pulse generation from the second
plateau of the HHG in Fig.2(red solid line). Due to the phase
mismatch, it is not suitable to select all the continuum spectrum of
the second plateau to synthesize single attosecond pulse. So we
impose a bandpass with bandwith of 90 order harmonics on the
continuum spectrum and make an inverse Fourier transformation, and
then an isolated 30 attoseconds pulse is obtained, as shown in
Fig.3(red line). In addition, we investigate the attosecond
generation from the laser pulses with different lengths. The HHG
spectrums for the pulse whose lengths are 70 O.C. and 80 O.C. are
shown in Fig.2. As displayed in this figure, though the pulse length
varies from 60 O.C. to 80 O.C., the HHG spectrum does not has
essential changes. All of the HHG spectrums have two plateaus and
the same cutoffs. This is easy to be understand from the laser field
in Fig.1(a). It is clear that the symmetry near the center of the
pulse is dramatically broken, while has not apparent changes in the
rest part of the laser field. That is to say, this type of nonlinear
chirp only cause the asymmetry of laser field near the center. Thus,
for multicycle pulses, the asymmetry has little differences when the
pulse length varies. So the HHG spectrum is almost independent of
the pulse length. The sub-50 attosceond pulses generated from the
HHG spectrums are shown in Fig.3, and the durations of these pulses
have not much difference, just as we expected.

\begin{figure}[htb]
\centerline{\includegraphics[width=8cm,height=4cm]{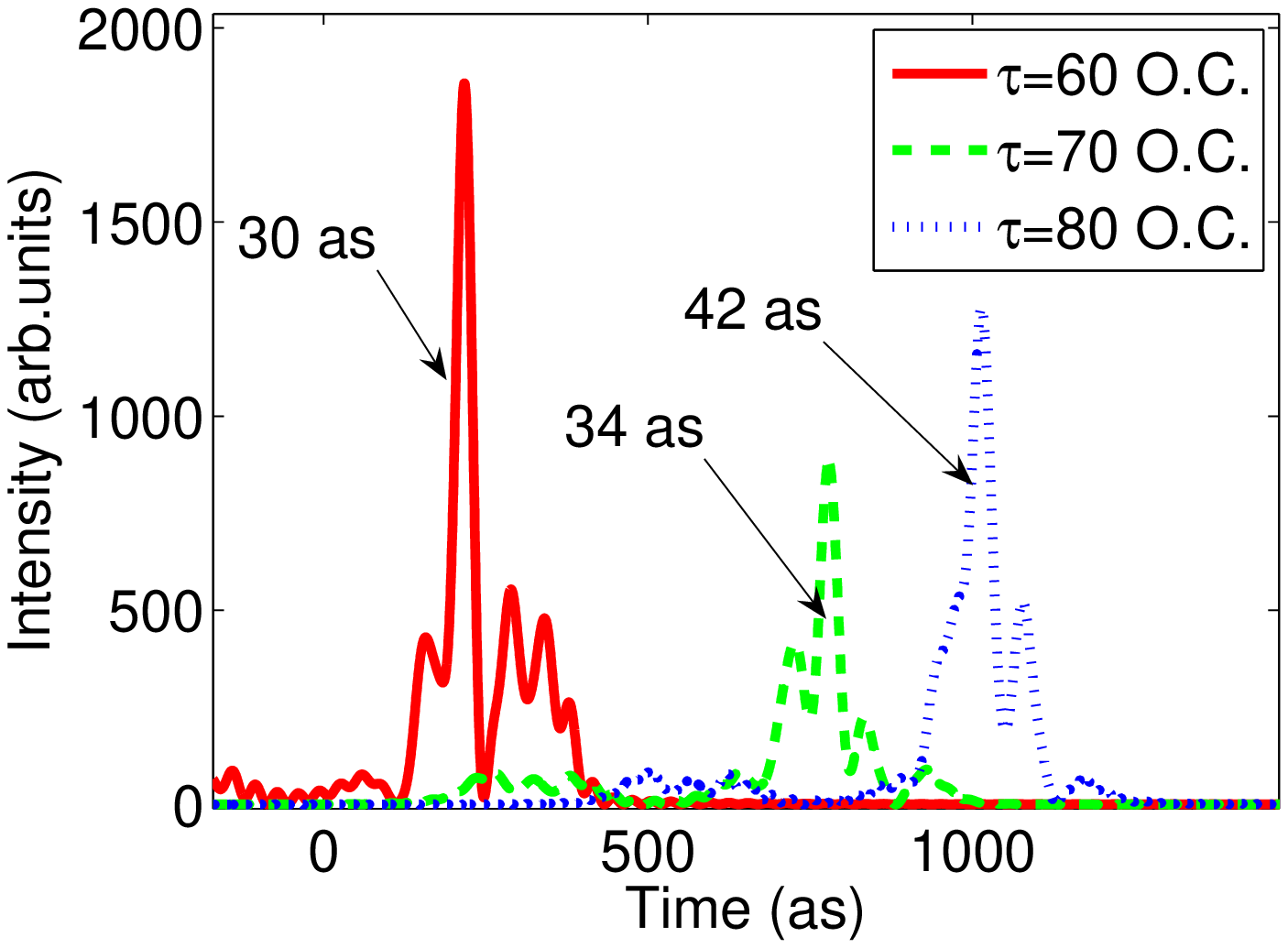}}
\caption{(Color online) The temporal profiles of the attosecond
pulses generated from the HHG spectums of Fig.2. The harmonic orders
used are all from the 275th to the 365th.}
\end{figure}

From the classical analysis above, the single attosecond generation
should not be much sensitive to the form of the chirp, if such
apparent asymmetry as displayed in Fig.1(a) can be obtained. To
verify this, we consider the other form of nonlinear chirps as:
\begin{align}
{{\delta}_{1}\left(t\right)} &=-8.0\arctan(\frac{t-50}{200}),
\end{align}
or
\begin{align}
{{\delta}_{2}\left(t\right)} &=-\frac{0.04t}{\sqrt{1+0.000015t^2}},
\end{align}
which also could be achieved by means of the comb laser
technology$^{11,12}$. The HHG spectrums for the two different
nonlinear chirped laser pulses are shown in Fig.4. Both the lengths
of the two pulses are 60 O.C.. From Fig.4, we can see that though
the forms of the chirps are quite different, the HHG are very
similar: both of them have two plateaus and the cutoffs energy are
almost the same. For comparison, the same order harmonics are used
to generate attosecond pulse, as shown in Fig.5. Clearly, the
durations of the pulses are below 50 attoseconds.
\begin{figure}[htb]
\centerline{\includegraphics[width=8cm,height=4cm]{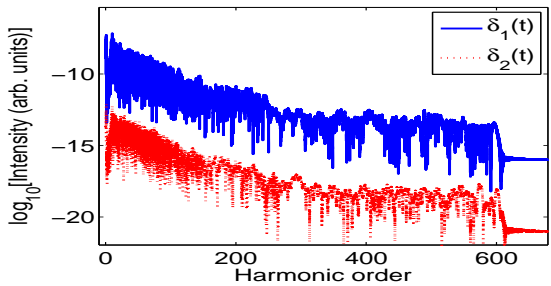}}
\caption{(Color online) The HHG spectra for the laser pulses with
different chirp. For clarity, the curves have been offset along the
vertical axis.}
\end{figure}
\begin{figure}[htb]
\centerline{\includegraphics[width=8cm,height=4cm]{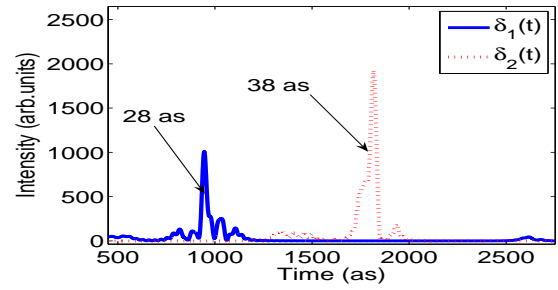}}
\caption{(Color online) The temporal profiles of the attosecond
pulses generated from the HHG spectra of Fig.4. The harmonic orders
used are both from the 275th to the 365th.}
\end{figure}

In conclusion, we proposed a method of using nonlinear chirp pulses
to generate a single attosecond pulse with duration less than 50
attoseconds in multicycle-driver regime. The results are almost
independent of the length and chirp form of the driver laser pulse,
which may make it easy to be realized in the present experiment
condition.

\bigskip
The work is supported by the National Basic Research Program of
China (Grant No.2006CB921104, 2006CB806000, 60708008, 10734080,
10874194), the Project of Academic Leaders in Shanghai (Grant
No.07XD14030) and the Key Basic Research Foundation of Shanghai
(Grant No. 08JC1409702).


\end{document}